\documentstyle[epsf,floats,preprint,pre,aps]{revtex}

\begin{document}

\draft 

\title{NONSTATIONARY OPTIMAL PATHS AND 
TAILS OF PREHISTORY PROBABILITY DENSITY IN
MULTISTABLE STOCHASTIC SYSTEMS }

\author{B. E. Vugmeister, J. Botina,   and H. Rabitz }
\address {\it Department of Chemistry, Princeton University, Princeton, NJ
 08544} 


\maketitle

\begin{abstract}

The tails of prehistory probability density in nonlinear multistable 
stochastic 
systems driven by white Gaussian noise, which has been a subject of
recent study, are analyzed  by   
employing the concepts of nonstationary optimal fluctuations.
 Results of numerical simulations evidence that  
the prehistory probability density is non-Gaussian and highly
asymmetrical that is an essential feature of noise driven fluctuations
in nonlinear systems. We show also that 
in systems with the detail balance the prehistory probability density is 
the conventional transition probability that obeys the backward Kolmogorov 
equation
\end{abstract}

\newpage

\section{Introduction}
 Recent theoretical and experimental studies\cite{D2,D3} of large
fluctuations in 
stochastic systems driven by white Gaussian noise have shown that 
among the different fluctuation paths of the system, the most probable 
{\it optimal} fluctuation path plays a crucial role. This is true    
for large output fluctuations, since the probability of encountering such 
fluctuations peaks sharply at the deterministic optimal fluctuation trajectory 
driven by an optimal realization from the random noise bath. 

The most probable fluctuation path, which we will call the stationary
optimal path (SOP), is the optimal trajectory $x_{opt}^s(t;t_f,x_f)$
that brings the system to a given point $x_f$ of the phase space at
instant $t_f$ from the vicinity of the initial attractor $x_{eq}$,
where the system has been fluctuating for a long period of time 
prior to reaching the point $x_f$.The concept of SOP can be traced back
to the work of Onsager-Machlup\cite{OM} and has been further widely used
 (see e.g., Ref.\cite{WF,D1,Marder,McKane,Maier,Naeh,Elber}).

 In particular, the SOP determines the quasistationary probability density
$P^{eq}(x_f)$, referenced to the local equilibrium point $x_{eq}$, that
system is located at point $x_f$ under the condition that it never
left the region of attraction to  $x_{eq}$.
Another quantity of interest is the transition probability density 
$P(x_f,t_f;x_0,t_0)$ for the system to
be at point $x_f$ at $t_f$ given that it was at  $x_0$ at time
$t_0$. It follows that $P(x_f,t_f;x_0,t_0) \rightarrow
P^{eq}(x_f)$ for $t_0 \rightarrow -\infty $ provided  $x_0$ belongs
to the domain of attraction to $x_{eq}$.

The distribution of different paths for a nonlinear double well
potential has been investigated\cite{D2,D3} through consideration of the so
called {\it prehistory} probability density $P_h(x_f,t_f;x_0,t_0)$ ,
that is a conditional probability that system will be brought to final
point $x_f$ at $t_f$ from the vicinity of the equilibrium position
$x_{eq}$ via the point $x_0$ at intermediate instant $t_0$.  Using the
fact that the prehistory probability density reaches its maximum on
the SOP ending at $x_f$, then $ \ln P_h(x_f,t_f;x_0,t_0)$ has been
represented as a power series with respect to the deviation of $x_0$
from the optimal path. The first term in this series is responsible
for the Gaussian behavior of the $P_h(x_f,t_f;x_0,t_0)$ near its
maximum. It is apparent that, the the validity of power series
expansion requires that the deviation of $x_0$ from the SOP be
sufficiently small.

Until now there were no attempts to calculate the prehistory
probability density outside of the Gaussian domain. The
region where one can observe the deviation of $P_h(x_f,t_f;x_0,t_0)$
from the universal Gaussian form is most interesting since it reflects
the specificity of the particular system.

Motivated by the fact that the deviation of $P_h(x_f,t_f;x_0,t_0)$
from the Gaussian form has been observed in  experiments\cite{D2},
in this paper we have performed an analysis of the prehistory
probability density for a broad range of initial positions $x_0$
outside the vicinity of the SOP. This approach is based on the calculation
of "nonstationary optimal paths", which we call the nonstationary
optimal trajectories that maximize the value of the transition
probability considered as a functional of different paths starting at
point $x_0$ at time $ t_0$ and ending at point $x_f$ at time $
t_f$. We will show below that the nonstationary optimal path formalism
naturally allows for calculation of  the highly non-Gaussian tails of the
prehistory probability density appearing in nonlinear systems.

\section{Nonstationary Optimal Paths in Nonlinear Systems} 

We will consider a one dimensional stochastic system  as it has been
the subject of modeling  in analog experiments\cite{D2}
described by the equation (dimensionless units) 

\begin{equation}
\dot{x} =- U'(x)  + f(t),    
\label{eq:ns1}
\end{equation}
where $U'(x)$ is the deterministic force induced by the nonlinear double 
well potential 

\begin{equation}
U(x) =-{1 \over 2} x^2 +{1 \over 4} x^4   
\label{eq:pot}
\end{equation}
and $f(t)$ is random Gaussian white noise with the correlation function  

\begin{equation}
\langle f(t) f(t') \rangle =D \delta (t-t').   
\label{eq:ns2}
\end{equation}

In order to calculate the
transition probabilities we make use of  Feynman's 
notion\cite{Feynman} of  relating  the probability functional 
$\Phi[x(t)]$ of the system output fluctuations and the probability
functional $P[f(t)]$ of the input noise.
For Gaussian noise the probability functional $P[f(t)]$ is given by 

\begin{equation}
P[f(t)] \propto \exp[-{1 \over 2 D} \int_{t_0}^{t_f} dt f(t)^2 ] 
\label{eq:ns3}
\end{equation}
One can see from Eq.(\ref{eq:ns3}) that $ P[f(t)]$, and hence
$\Phi[x(t)]$, reaches its maximum for the most probable {\it optimal}
random field $f_{opt}$, introduced in Ref.\cite{D1}, which minimizes 
the integral $\int dt f(t)^2$
under the constraint that the equation of motion , Eq.(\ref{eq:ns1}),
is satisfied and $x(t_0)=x_0, x(t_f) =x_f$. Such a constraint leads to
a nonzero value of the optimal noise field $f_{opt}(t) $ which would be zero
without constraints. We will assume that points $x_0$ and $x_f$ belong
to the region of attraction of the same attractor including any small
vicinity of the separatrix.

The optimal transition probability is given by the most
favorable realization of noise which  brings the system from point
$(x_0,t_0)$ to point $(x_f,t_f)$, i.e., 

\begin{equation}
P(x_f,t_f;x_0,t_0) \propto \exp[-{1 \over 2D} 
\int_{t_0}^{t_f} dt f_{opt}(t)^2 ] 
\label{eq:ns6}
\end{equation}
In order to find the  quasistationary probability distribution
$P^{eq}(x_f)$ one should take the  limit in
Eq.(\ref{eq:ns6}) $x_0 \rightarrow x_{eq}, t_0 \rightarrow-\infty $.
The corresponding optimal trajectory is a  SOP and the corresponding
optimal field is a quasistationary optimal field $f_{opt}^s(t)$. 
We have 

\begin{equation}
P^{eq}(x_f) \propto \exp[-{1 \over 2D} 
\int_{-\infty}^{t_f} dt f_{opt}^s(t)^2 ] 
\label{eq:ns6a}
\end{equation}

Following Ref.\cite{D1} we replace $f(t)$ in Eq.(\ref{eq:ns6}) by its
form given by the equation of motion, Eq.(\ref{eq:ns1}). We obtain 

\begin{equation}
P(x_f,t_f;x_0,t_0) \propto \exp[-{S \over 2D}]
 \label{eq:P}
\end{equation}
where S is the action integral

\begin{equation}
S={1 \over 2} \int_{t_0}^{t_f} dt(\dot{x} +U'(x))^2
\label{eq:S}
\end{equation}

Note that in the case of white noise Eq.(\ref{eq:P}) can be obtained
also from the corresponding Fokker-Plank equation with the use of WKB
approximation\cite{Ben}. The validity of Eqs.(\ref{eq:ns6}),(\ref{eq:S})
corresponds to the limit of low noise intensity for which ${S\over D} >>
1$.

Variation  of the resulting action integral gives rise to effective dynamics 
described by the Hamiltonian

\begin{equation}
H={1 \over 2} \dot{x}^2 - {1 \over 2} U'(x)^2
\label{eq:H}
\end{equation}
and equation of motion

\begin{equation}
\ddot{x} -U'(x) U''(x) =0
\label{eq:ns7}
\end{equation}
with boundary condition 

 \begin{equation}
x(t_0) =x_0;~~~~~x(t_f)=x_f
\label{eq:ns8}
\end{equation} 
Eq.(\ref{eq:ns7}) is the Euler-Lagrange equation for the
functional  Eq.(\ref{eq:S}).

Solution of Eqs.(\ref{eq:ns7}),(\ref{eq:ns8}) represents the nonstationary
optimal path $x_{opt}(t)$ between points $(x_0,t_0)$ and $(x_f,t_f)$
corresponding to finite energy in Eq.(\ref{eq:H}). The 
SOP  $x_{opt}^s(t;x_f,t_f)$ ending at point $x_f$ at $t_f$ 
represents a partial solution of Eq.(\ref{eq:ns7}) that satisfies the first 
order differential equation

\begin{equation}
{d x_{opt}^s \over dt} = U'(x_{opt}^s)
\label{eq:ns9}
\end{equation}
with boundary conditions

\begin{equation}
x(t_f) =x_f;~~~~~x(-\infty)=x_{eq}
\label{eq:ns10}
\end{equation}
The solution of Eqs.(\ref{eq:ns9}) and (\ref{eq:ns10}) with
$x_{eq}=-1$ for  the double well 
potential given by Eq.(\ref{eq:pot}) is the {\it instanton} 
solution\cite{Raj}

\begin{equation}
x_{opt}^s(t) = -{1 \over \sqrt{1+C \exp[ 2 t]}}, ~~~~~
f_{opt}^s(t) = {2 C \exp[2t] \over \sqrt{(1+C \exp[ 2 t])^{3}}}  
\label{eq:inst}
\end{equation}
where the constant C determines the value $x_{opt}^s(t_f) =x_f$.  The
other partial solution of the second order differential equation
Eq.(\ref{eq:ns7}) satisfies the equation ${d x\over dt}=-U'(x)$, that
is the equation of motion, Eq.(\ref{eq:ns7}) without noise. Note that
in the general case with a nonlinear dependence of $U'(x)$ on $x$, the
solution of Eq.(\ref{eq:ns7}) can not be presented as a linear
combination of the partial solutions.

The nonstationary optimal paths for the potential given by
Eq.(\ref{eq:pot}) have been found by numerical integration of
Eq.(\ref{eq:ns7}) subject to boundary conditions,
Eq.(\ref{eq:ns8}). The two point boundary value problem has been
reexpressed by considering the problem with initial conditions $x(t_0)
=x_0,~ \dot{x}(t_0) =v_0$. A minimization procedure has been employed
in order to find the best value of the initial velocity $v_0$ to reach
the target point $x_f $ at $t_f$. As a test for the calculations the
energy conservation law in Eq.(\ref{eq:H}) has been verified for each
trajectory obtained.
 
In Figs.1a and 2a we illustrate the typical nonstationary optimal
paths corresponding to $x_f=-0.1$ for  different values of $x_0$
and $\tau= t_f -t_0$.  For comparison we show also the SOP, given by
Eq.(\ref{eq:inst}) (for $x_f=-0.1$ the constant C=99 in
Eq.(\ref{eq:inst}) ).  As one can see from Fig.2a, the essential
feature of the nonstationary optimal trajectories starting at $x_0 <
0$ is the possibility of a sign  change of the velocity $\dot{x}(t)$ 
at an intermediate point of the trajectory.  For small values of
$\tau$ nonstationary optimal trajectories deviate significantly from
the SOP, whereas for asymptotically large $\tau$ they rapidly approach
the SOP independently of the initial point $x_0$. This behavior  can 
also be understood from the exact solution of Eq.(\ref{eq:ns7})
for the harmonic potential presented in the next section.

With the use of Eq.(\ref{eq:ns1}) and the known temporal form of the
nonstationary trajectories one can reproduce the corresponding values
of the optimal noise field . The values of $f_{opt}(t)$ for the
trajectories shown in Figs.1a and 2a are presented in Figs.1b and 2b.

In the next section the approach above will be used for the calculation of 
the prehistory probability density for a broad range of initial positions 
$x_0$.

\section{Prehistory Probability Density}

The prehistory probability density  is given by\cite{D2}

\begin{equation}
P_h(x_f,t_f;x_0,t_0) \propto \exp[-{1 \over D} 
(\int_{-\infty}^{t_f} dt f_{opt}(t,x_0,x_f)^2 - 
\int_{-\infty}^{t_f} dt f_{opt}^s(t,x_f)^2 )  ] 
\label{eq:ns4}
\end{equation}
where the optimal noise field $f(t,x_0,x_f)$ induces the nonstationary
optimal trajectory which  starts at $x_{eq}$ at $t=-\infty$, passes
point $ x_0 $ at $t_0$ and reaches point $x_f$ at $t_f$; $f_{opt}^s$
is the stationary optimal field that induces the SOP ending at point $x_f$
at $t_f$. Thus we may write 

\begin{equation}
\int_{-\infty}^{t_f} dt f_{opt}(t,x_0,x_f)^2 =
\int_{-\infty}^{t_0} dt f_{opt}^s(t, x_0)^2 + 
\int_{t_0}^{t_f} dt f_{opt}(t,x_0,x_f)^2,
\label{eq:int1}
\end{equation}
and with  Eq.(\ref{eq:ns4}), Eq.(\ref{eq:int1}) we arrive at  

\begin{equation}
P_h(x_f,t_f;x_0,t_0) ={P^{eq}(x_0) P(x_f,t_f;x_0,t_0)\over P^{eq}(x_f)}  
\label{eq:ns5}
\end{equation}

The difference between $P_h(x_f,t_f;x_0,t_0)$  and $P(x_f,t_f;x_0,t_0)$ 
is that 
$P_h(x_f,t_f;x_0,t_0)$ satisfies the normalization  condition with respect to 
initial points $x_0$ ($\int dx_0 P_h(x_f,t_f;x_0,t_0) =1 $) whereas 
$P(x_f,t_f;x_0,t_0)$  satisfies the analogous normalization  condition 
with respect to final points $x_f$. Eq.(\ref{eq:ns4}) shows that if
$x_0$ belongs to 
the stationary optimal path ending at $x_f$, then both integrals in  
Eq.(\ref{eq:ns4}) are equal and cancel meaning that 
$ P_h(x_f,t_f;x_0,t_0)$ reaches its maximum on the stationary optimal
path \cite{D2}. 

It follows from  Eq.(\ref{eq:ns5}) that 
$P^{eq}(x_f)$ given by 
Eqs.(\ref{eq:ns6a}),(\ref{eq:ns1}),(\ref{eq:inst}) satisfies the
principal of detailed balance  

\begin{equation}
{ P^{eq}(x_f) \over P^{eq}(x_0)} =\exp[-{1 \over D} (U(x_f) -U(x_0))]  
\end{equation}
and    
$P(x_f,t_f;x_0,t_0)$ is the conventional transition probability that
obeys the forward Kolmogorov equation\cite{Gard}
due to the white character of the noise.  Then the prehistory probability
density should obey the backward Kolmogorov equation and, in fact, does 
not depend on the prehistory. Below we will illustrate this conclusion,  
which is a consequence of  the chosen quasiequilibrium value of the 
initial distribution, on the model system with a quasiharmonic potential.
In the general case the prehistory probability density does depend on
prehistory in multistable stochastic systems. 

Note also that being consistent with the concept of optimal fluctuation
we assume that $U(x_f) > U(x_0) $ ( a particle is ``climbing uphill'')
resulting in occasional fluctuations described by the prehistory
probability density.

In Refs.\cite{D2,D3} the partial solution, Eq.(\ref{eq:ns9}) has been used
for the evaluation of the prehistory probability density based on the 
iterative  solution of Eq.(\ref{eq:ns7}) near the SOP  implying  
that the values of $(x_0 -x_{opt}^s(t_0; x_f,t_f))$ are not too large. 
We stress, however, that using the formalism above, the prehistory
probability density as well as the transition 
probabilities can be evaluated  without the 
limitation on the values of $(x_0 -x_{opt}^s(t_0; x_f,t_f))$.

Prior to concentrating  on  the results of numerical calculations for the
case of the nonlinear potential, Eq.(\ref{eq:pot}), we will present
for illustration the simple  expressions for
$P(x_f,t_f;x_0,t_0)$ and $P_h(x_f,t_f;x_0,t_0)$ with the
quasiharmonic potential of the form

\begin{equation}
U(x)= \cases{
{1 \over 2} x^2,  &$ x \leq 1 $\cr
{1\over 2}(x-2)^2, &$  x  > 1$ \cr}
\label{eq:a13}
\end{equation}
In this case the solution of Eq.(\ref{eq:ns7}) can be obtained analytically.
We have in the domain of attraction at  $x_{eq}=0$,  

\begin{equation}
x_{opt}(t)= {1 \over e^{t_f -t_0} - e^{t_0 -t_f}} [ 
e^t ( x_f e^{-t_0} -x_0 e^{-t_f}) + e^{-t} ( x_0 e^{t_f} -x_f
e^{t_0})],~~ x < 1
\label{eq:ns11}
\end{equation}
where $x_{opt}(t)$ is the nonstationary optimal path.

The optimal noise field is of the form  

\begin{equation}
f_{opt}(t)= {2 e^t \over e^{t_f -t_0} - e^{t_0 -t_f}}  
 ( x_f e^{-t_0} -x_0 e^{-t_f}) 
\end{equation}

Note that the two terms in Eq.(\ref{eq:ns11}) proportional to $e^t$
and $e^{-t}$ 
respectively represent the two partial solutions of
Eq.(\ref{eq:ns7}). Only the  partial solution  proportional to $e^t$,
which reaches a finite limit at  $t_0 \rightarrow -\infty$,
contributes to the SOP. In fact,  
for  $t_0 \rightarrow -\infty$, it follows from Eq.(\ref{eq:ns11})
that  $x_{opt}(t)=x_{opt}^s(t; x_f,t_f)=x_f e^{t-t_f}$\cite{OM}.  

We obtain from Eqs.(\ref{eq:ns6}) and  (\ref{eq:ns5}),

\begin{equation}
P(x_f,t_f;x_0,t_0) \propto \exp[- {1 \over D} {(x_f -x_0 e^{-(t_f-t_0)})^2
\over 1-e^{-2 (t_f -t_0)}}]
\label{eq:ns13}
\end{equation}

 \begin{equation}
P_h(x_f,t_f;x_0,t_0) \propto \exp[- {1 \over D} {(x_0 -x_f e^{-(t_f-t_0)})^2
\over 1-e^{-2 (t_f -t_0)}}]
\label{eq:ns14}
\end{equation}
One can see from  Eqs.(\ref{eq:ns13}) and (\ref{eq:ns14}) that in the case of 
the quasiharmonic potential the nonstationary optimal path approach 
reproduces the exact results for the transition probability and the
prehistory probability density  
in linear stochastic systems with Gaussian noise. Note that the 
prehistory probability density given by Eq.(\ref{eq:ns14}) obeys the 
backward  Kolmogorov equation\cite{Gard}. 

The prehistory probability density for the nonlinear potential given by 
Eq.(\ref{eq:pot}) has been calculated numerically with the use of 
Eq.(\ref{eq:ns4}). Precaution is required, however, since near the SOP 
the exponent  in Eq.(\ref{eq:ns4}) becomes the small difference of large 
numbers. In order to improve the accuracy of the
calculations we may represent the exponent of 
$P_h(x_f,t_f;x_0,t_0)$ in the form 

\begin{equation}
\int_{-\infty}^{t_f} dt[ f_{opt}^s(t,x_0)^2 - f_{opt}^s(t, x_f)^2] + 
\int_{t_0}^{t_f} dt [f_{opt}(t,x_0,x_f)^2 - f_{opt}^s(t,x_f)^2]
\label{eq:int2}
\end{equation}
One can see from Eq.(\ref{eq:int2}) that if $x_0$ belongs to the
stationary optimal path ending at point $x_f$ at $t_f$, then both
integrands in Eq.(\ref{eq:int2} vanish.  The calculated values of $- D
\ln P_h(-0.1,0;x_0,-\tau)$ are presented in Fig.3. It is seen that the
parabolic character of the curves, and hence the Gaussian character of
prehistory probability density, takes place only in a small region
in the vicinity of SOP where the prehistory probability density
reaches its maximum. In general, the function $P_h(x_f,t_f;x_0,t_0)$
is highly asymmetrical, and that is the essential feature of the path
distribution in nonlinear systems.

\section{Conclusion}

We have shown that the concept of a nonstationary optimal path is an
adequate approach for the analysis of transition probabilities and
prehistory probability density in noise driven systems. The concept
allows one to analyze the optimal path distribution outside of the immediate
vicinity of the stationary optimal path. The observed highly
asymmetrical shape of the prehistory probability density is the
essential feature of the fluctuations in nonlinear noise driven
systems. We hope that this observation will stimulate additional
experiments on the analysis of optimal path distributions in nonlinear
systems.

The nonstationary optimal path approach for the analysis of fluctuations 
in stochastic systems should be especially useful for
exploring the  possibilities of  control of fluctuations by an external
field. It has  been shown recently\cite{VR,SD} that an optimal
control field with a finite time duration can naturally cooperate with the
nonstationary optimal noise such  that its temporal form coincides with
the temporal form of the optimal fluctuations.

\acknowledgments

We are thankful to M.I. Dykman, R.S. Maier, M. Marder and
V.N. Smelyanskiy for the useful discussions of the results of the
paper. The authors acknowledge support from the Office of Naval
Research and National Science Foundation.

\newpage

\begin{figure}

\caption{ The nonstationary optimal path (a) and optimal field (b)
for the nonlinear potential Eq.(\ref{eq:pot}) and $\tau =2 $ corresponding
to the final point $x_f=-0.1$ and initial points $x_0 = -1~ (1);~ -1.5 ~(2);~ -0.5~ (3).$
The dotted lines show the stationary optimal path and the corresponding
optimal field. }
\vskip 1truecm

\caption{The nonstationary optimal path (a) and optimal field(b)
for the nonlinear potential Eq.(\ref{eq:pot}) and $\tau =8 $ corresponding
to the final point $x_f=-0.1$ and initial points $x_0 = -1~ (1);~ -1.5 ~(2);~ -0.5 ~(3).$
The dotted lines show the stationary optimal path and the corresponding
optimal field. } 

\vskip 1truecm 

\caption {The activation energy of the prehistory probability density for
$x_f=-0.1$ as a function of initial position $x_0;~ 1)-\tau=8,~ 2)
\tau =2,~ 3)\tau=1.5,  ~4)\tau= 1 ,~ 5)\tau= 0.5. $ The data for curves
2-5 are multiplied by 10. Stars represent the values of $2 (U(x_0-U(x_{eq}) )$
and confirm that quasiequilibrium distribution takes place for large $\tau$.}
\end{figure}
\newpage

\hskip -2truecm 
\begin{figure}
\epsffile{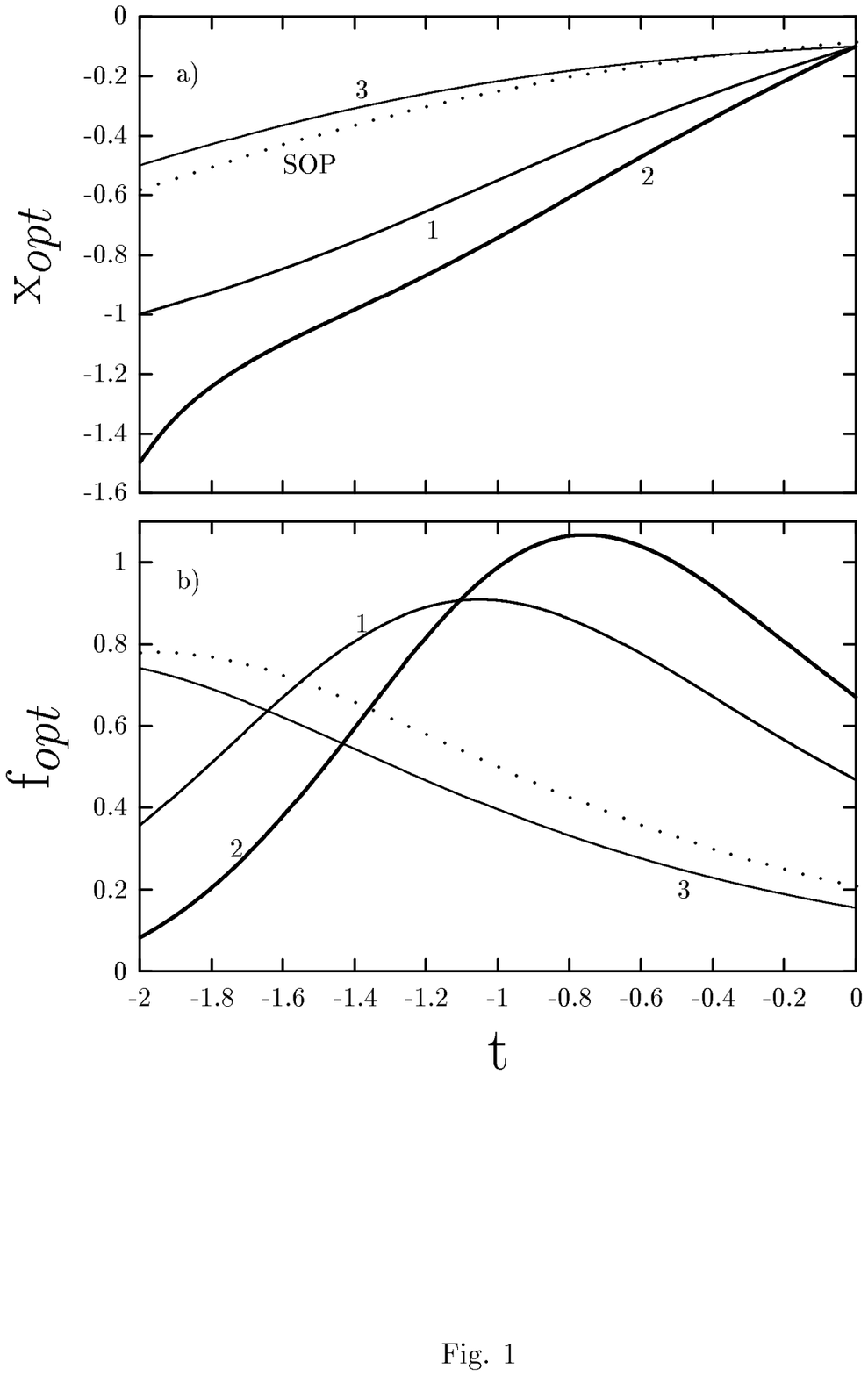}
\end{figure}
\newpage
\newpage
\begin{figure}
\epsffile{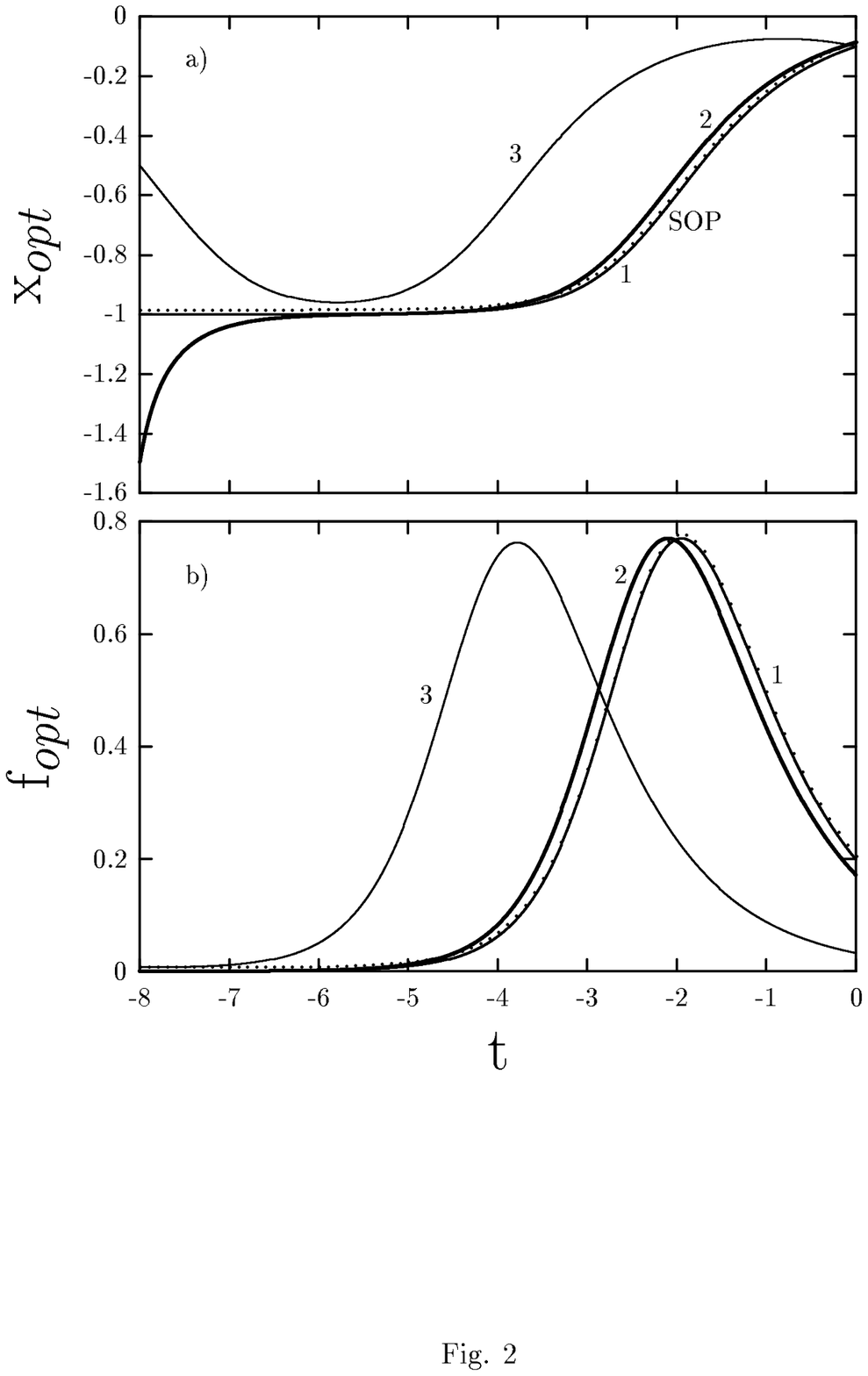}
\end{figure}
\newpage
\begin{figure}
\epsfxsize=7in
\epsffile{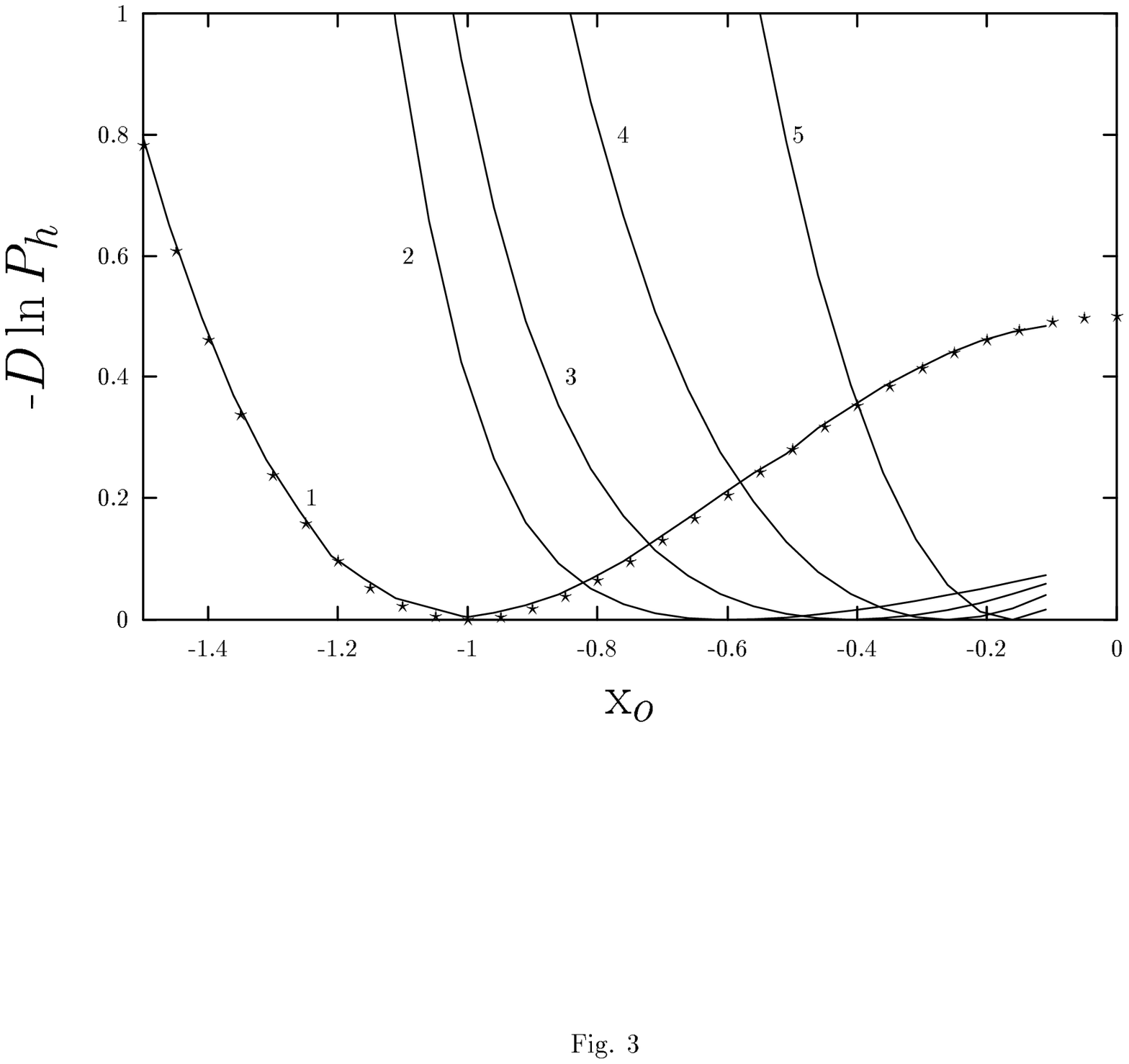}
\end{figure}


\begin{thebibliography}{99}




\bibitem{D2} M.I. Dykman, P.V.E. McClintock, V.N. Smelyanskiy,
N.D. Stein, and N.G. Stocks, Phys. Rev. Lett. {\bf 68}, 2718 (1992).

\bibitem{D3} M.I. Dykman, D.G. Luchinckiy, P.V.E. McClintock, and
V.N. Smelyanskiy, Phys. Rev. Lett. 1996.

\bibitem{OM} L. Onsager and S.Machlup, Phys.Rev. {\bf 91},1505, 1512 (1953).

\bibitem{WF} A.~D. Wentzell and M.~I. Freidlin, {\it Russ.\ Math.\
Surveys\/} {\bf 25}(1) (1970); M.~I. Freidlin and A.~D. Wentzell, {\it
Random Perturbations of Dynamical Systems} (Springer-Verlag, New
York/Berlin, 1984).


\bibitem{D1} M.I. Dykman and M.A. Krivoglaz, Sov.Phys. JETP {\bf 50},
30(1979); M.I. Dykman, Phys. Rev. A {\bf 42}, 2020 (1990);

\bibitem{Marder} M. Marder Phys.Rev.Lett. {\bf 74}, 4547 (1995);
Phys. Rev.E {\bf 54},3442 (1996).

\bibitem{McKane} A.J.Bray and A.J. McKane, Phys.Rev.Lett. {\bf 62},
493 (1987). 

\bibitem{Maier} R.S.Maier and D.L.Stein, Phys. Rev.Lett. {\bf 69},3691
(1992); {\bf  71}, 1783 (1993); {\bf 77},4861 (1996);
J. Stat. Phys. {\bf 83},291 (1996) 

\bibitem{Naeh} T. Naeh, M.M. Klosek, B.J.Matkowsky, and Z. Schuss,
SIAM J.Appl. Math. {\bf 50}, 595(1990).

\bibitem{Elber} R. Olender and R. Elber, J. Chem. Phys. {\bf 105},9299
(1996). 

\bibitem{Feynman}  R.P. Feynman and A.R. Hibbs, {\it Quantum Mechanics
and Path Integrals}  (McGraw-Hill, NY 1965).

\bibitem{Ben} E. Ben-Jacob, D.J. Bergman, B.J. Matkowsky, and
Z. Schuss, Phys. Rev. A {\bf 26}, 2805 (1982)
 
\bibitem{Raj} R. Rajaraman, {\it Solitons and Instantons}
(North-Holland, Amsterdam, 1982).


\bibitem{Gard} G. Gardiner,{\it Handbook of Stochastic methods } 
(Springer-Verlag, 1985).


\bibitem{VR} B.E. Vugmeister and H. Rabitz, Phys. Rev. E {\bf 55} No 3 (1997).

\bibitem{SD} V.N. Smelyanskiy and M.I.Dykman, Phys. Rev. E {\bf 55} No 3 (1997).


\end{thebibliography}
\end{document}